\documentclass[twocolumn, conference]{IEEEtran}
\usepackage{amssymb}
\usepackage{amsmath}
\usepackage{amsfonts,amsthm}
\usepackage{cite}
\usepackage{color}
\usepackage{url}
\usepackage{graphicx}
\usepackage{yfonts}

\IEEEoverridecommandlockouts

\DeclareMathOperator{\E}{\mathbb{E}}

\newcommand{\B}[1]{{\pmb{#1}}}

\newcommand{\CG}[2]{\mathcal{CN}\left({#1},{#2}\right)}

\newcommand{\Source}[1]{{S}_{#1}}
\newcommand{\Destination}[1]{{D}_{#1}}
\newcommand{\Relay}{{R}}
\newcommand{\nR}{N}

\newcommand{\yR}{\B{y}_{{R}}}

\newcommand{\nRe}{\B{n}_{{R}}}

\newcommand{\np}{\sigma^2_{n}}

\newcommand{\yDk}{y_{{D}_k}}
\newcommand{\nDk}{n_{{D}_k}}

\begin{document}

\title{Multi-Pair Amplify-and-Forward Relaying with Very Large Antenna Arrays}

\author{\authorblockN{Himal A. Suraweera\authorrefmark{1}, Hien Quoc Ngo\authorrefmark{2}, Trung Q. Duong\authorrefmark{3}, Chau Yuen\authorrefmark{1} and Erik G. Larsson\authorrefmark{2}}
\authorblockA{\authorrefmark{1}\footnotesize Singapore University of Technology and Design, Singapore (e-mail: \{himalsuraweera, yuenchau\}@sutd.edu.sg)}
\authorblockA{\authorrefmark{2}\footnotesize Department of Electrical Engineering (ISY), Link\"{o}ping University, Link\"{o}ping,  Sweden (e-mail: \{nqhien, egl\}@isy.liu.se)}
\authorblockA{\authorrefmark{3}\footnotesize Blekinge Institute of Technology, Karlskrona, Sweden (e-mail: quang.trung.duong@bth.se)}
\thanks{This research is partly supported by the Singapore University Technology and Design (Grant No. SUTD-ZJU/RES/02/2011). The work of H. Q. Ngo and E. G. Larsson was supported  in part by the Swedish Research Council (VR), the Swedish Foundation for Strategic Research (SSF), and ELLIIT.}}

\maketitle
%
\begin{abstract}
We consider a multi-pair relay channel where multiple sources simultaneously communicate with destinations using a relay. Each source or destination has only a single antenna, while the relay is equipped with a very large antenna array. We investigate the power efficiency of this system when maximum ratio combining/maximal ratio transmission (MRC/MRT) or zero-forcing (ZF) processing is used at the relay. Using a very large array, the transmit power of each source or relay (or both) can be made inversely proportional to the number of relay antennas while maintaining a given quality-of-service. At the same time, the achievable sum rate can be increased by a factor of the number of source-destination pairs. We show that when the number of antennas grows to infinity, the asymptotic achievable rates of MRC/MRT and ZF are the same if we scale the power at the sources. Depending on the large scale fading effect, MRC/MRT can outperform ZF or vice versa if we scale the power at the relay.
\end{abstract}


\section{Introduction}
Multiple-input multiple-output (MIMO) technology has now become an
integral feature of many advanced communication systems. A
cellular architecture with MIMO that has gained significant
research interest is multi-user MIMO (MU-MIMO) in which an antenna
array simultaneously serves a multiplicity of autonomous
co-channel users/nodes \cite{GKHCS:07:SPM}. While the current
systems have limited number of antennas (e.g. the LTE standard
allows for up to $8$ antenna ports), MU systems having a large
number of antennas at the base station (\emph{very large MIMO})
have been advocated recently in
\cite{TLM2010TWC,RPLLMET:11:SPM,shepard}. Very large MU-MIMO
systems can substantially reduce the interference with simple
signal processing techniques and achieve increased reliability and
throughput, and significant reduction in total
transmitted power \cite{Ngo}.

On the other hand, relaying has been extensively explored to
provide expanded coverage and high throughput, especially at the
cell edge \cite{dohler}. However, inter-user interference can
cause major performance degradation in MU relay systems
\cite{AA2008TWC}. As a result, a large body of performance
analysis work on MU relay systems, e.g.,
\cite{NY2010TVT,HD2011TVT,JK2011TWC} has mainly avoided
interference slots by adopting spectrally inefficient policies
such as orthogonalization of time/frequency. Another line of work
on MU relay systems has considered deploying complicated
precoder/decoder designs; e.g., in \cite{CAO} and advanced
joint network coding and signal alignment techniques for the multi-pair two-way relay channel \cite{ZHAO}.

In this paper, we analyze the performance of a multi-pair relaying
scenario where a group of $K$ sources and $K$ destinations
communicate using a single relay equipped with $N$ antennas, where
$N \gg K$. To the best of our knowledge, there is no prior work
that analyzes the effects of large antenna arrays on the performance of
the considered relay system. In the first time slot, all $K$
sources simultaneously transmit their signals to the relay. In the second time
slot, a linearly transformed version of the received signal at the
relay is forwarded to the $K$ destinations. For this multi-pair
relay channel, we study the achievable rate vs. power efficiency
performance with (1) maximum ratio combining/maximal ratio
transmission (MRC/MRT) and (2) zero-forcing (ZF) at the relay.


We show that when $N$ is large, we can
cut the transmit power at each source or/and the relay
proportionally to $1/N$ with no performance degradation. The asymptotic achievable rates of MRC/MRT and ZF for $N \to \infty$ are
derived for cases when the transmit power of each source
or/and the relay is made inversely proportional to $N$. The
results show that when the transmit power of each
source scales as $1/N$ while keeping a fixed transmit power at the
relay, the fast fading, interference from other sources, and noise
at the destination disappear and hence, the system performance
does not depend on the quality of the channel in the second hop. In
contrast, for the case when the transmit power of each source is
fixed and the transmit power of the relay is scaled down as $1/N$, the
system performance does not depend on the channel quality of the
first hop. We further show that when the transmit power at the relay
is cut proportionally to $1/N$, with very large $N$, depending on
the large-scale fading effect, MRC/MRT performs better than ZF or
vice versa.

\emph{Notation}: $\dagger$, $||\cdot||$ and $\mathsf{Tr}\left(\cdot\right)$ denote the
conjugate transpose operation, Euclidean norm and the trace of a matrix respectively. $\E\{x\}$ stands for the expectation of a random variable $x$, and $\B{I}_M$ is the identity matrix of size $M$. $\mathop  \to \limits^{a.s.} $ and $\mathop  \to \limits^{d} $ denote the almost sure
convergence and  the convergence in distribution, respectively.

\begin{figure}[t]
\centering
\includegraphics[width=0.8\linewidth]{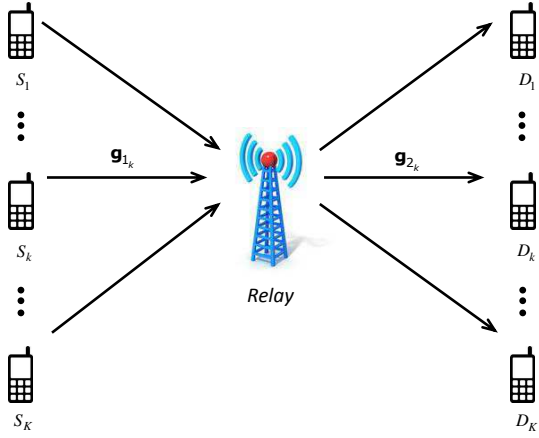}
\caption{System model. The channel strengths of the source-relay,
($\Source{k}-\Relay$) and relay-destination,
($\Relay-\Destination{k}$) links for $k=1,\ldots,K$ are
$\eta_{1k}$ and $\eta_{2k}$, respectively. }\label{fig:fig2}
\end{figure}
\section{System Model}
Consider a scenario with a group of $K$ sources, $\Source{k}$, $K$
destinations, $\Destination{k}$, for $k=1,\dots,K$, and a single
relay, $\Relay$. Each source/destination is equipped with a single
antenna while $\Relay$ is equipped with $\nR$ antennas as shown in
Fig. 1. The source $\Source{k}$ wants to communicate with the
destination $\Destination{k}$. Communication in this network
occurs via $\Relay$ since there are no direct links among
$\Source{k}$ and $\Destination{k}$, for any $k$ due to heavy
shadowing and path loss phenomenon.

During the first phase all sources simultaneously transmit their
symbols to $\Relay$, and the received $\nR \times 1$ signal vector
can be written as
\begin{align}
    \yR = \sqrt{P_\mathsf{t}}\B{G}_1 \B{x} + \nRe,
\end{align}
where
$\sqrt{P_\mathsf{t}}\B{x}=\sqrt{P_\mathsf{t}}[x_1,x_2,\dots,x_K]^{T}$
are transmitted symbols with $\E\{\B{x}\B{x}^\dagger\}=\B{I}_{K}$
(the average transmitted power of each source is $P_\mathsf{t}$)
and $\nRe$ is an $\nR \times 1$ additive white Gaussian noise
(AWGN) vector at the relay node with
$\E\{\B{n}_{\Relay}\B{n}_{\Relay}^\dagger\}=\np\B{I}_{\nR}$. The
$\nR \times K$ channel matrix between the $K$ sources and $\Relay$ is
expressed as $\B{G}_1=\B{H}_1\B{D}_1^{1/2}$ where $\B{H}_1$
contains independent and identically distributed (i.i.d.)
$\CG{0}{1}$ entries and $\B{D}_1$ is a $K \times K$ diagonal
matrix, where $[\B{D}_1]_{kk}=\eta_{1k}$. Moreover, we model the
$\nR \times K$ channel matrix between the $K$ destinations and
$\Relay$ as $\B{G}_2=\B{H}_2\B{D}_2^{1/2}$ where $\B{H}_2$
contains i.i.d. $\CG{0}{1}$ entries and $\B{D}_2$ is a $K \times K$
diagonal matrix, where $[\B{D}_2]_{kk}=\eta_{2k}$. Note that
$\B{H}_1$ and $\B{H}_2$ represent independent fast fading, while
$\B{D}_1$ and $\B{D}_2$ represent path-loss attenuation, and
log-normal shadow fading. The assumption of independent fast
fading is sufficiently realistic for systems where the antennas are
spaced sufficiently far apart \cite{Ngo}.

During the second phase, $\Relay$ re-transmits a transformation of
the received signal given by $\widetilde{\B{y}}_R=\B{W}\yR$. The signal at $\Destination{k}$ can be expressed as
\begin{align}
\label{sigr}
   \yDk = \sqrt{P_\mathsf{t}} \B{g}_{2_k}^{\dag} \B{W} \B{g}_{1_k} x_k
    &+\sqrt{P_\mathsf{t}}\sum_{i=1,i\neq k}^{K}
    \B{g}_{2_k}^{\dag} \B{W} \B{g}_{1_i} x_i
        \nonumber
    \\
    &+ \B{g}_{2_k}^{\dag} \B{W} \nRe + \nDk,
\end{align}
where $\B{g}_{1_i}$ is the $i$-th column of $\B{G}_1$,
$\B{g}_{2_k}$ is the $k$-th column of $\B{G}_2$, $\B{W}$ is the
$\nR \times \nR$ transformation matrix normalized to satisfy a
total power constraint, $P_{\mathsf{r}}$, at the relay as
$\mathsf{Tr}\left(\mathtt{E}\left\{\widetilde{\B{y}}_R\widetilde{\B{y}}_R^{\dag}\right\}\right)=P_{\mathsf{r}}$,
and $\nDk$ is the AWGN at $\Destination{k}$ with $\mathtt{E}\{\nDk
\nDk^\dagger\}=\np$.

As a result, the instantaneous end-to-end (e2e) signal-to-interference-noise ratio (SINR) at $\Destination{k}$ can be written as
\begin{align} \label{eq:SINR:k}
    \gamma_{k}
      =
    \frac{P_\mathsf{t}|\B{g}_{2_k}^{\dag} \B{W} \B{g}_{1_k}|^2}
    {P_\mathsf{t}\sum\limits_{i=1,i\neq k}^{K} |\B{g}_{2_k}^{\dag} \B{W} \B{g}_{1_i}|^2
    +\|\B{g}_{2_k}^{\dag} \B{W}\|^2 \np +\np
    }.
\end{align}

\subsection{MRC/MRT at the Relay}
When CSI is available at $\Relay$, it is natural to apply a
transformation based on the MRC/MRT principle\footnote{Note that
the choice of $\B{W}$ based on the MRC/MRT principle is not
optimal for maximizing the SINR. However, finding the optimal
$\B{W}$ in an analytical form seems impossible due to the
non-convex nature of the problem. Furthermore, with very large
antennas arrays, the channel vectors are nearly orthogonal, and
hence MRC/MRT is nearly optimal \cite{TLM2010TWC}.}. In the first
phase, the relay uses MRC to combine the signals transmitted from
$K$ sources and then in the second phase, it uses MRT precoding to
forward data to $K$ destinations. Hence the relay transformation
matrix is given by $\B{W}=a_{\mathsf{mrc}} \B{G}_2\B{G}_1^{\dag}$.
In this case, to meet the power constraint at the relay, we have
\begin{align}
a_{\mathsf{mrc}} =
\sqrt{\frac{P_{\mathsf{r}}}{\mathsf{Tr}\left(P_{\mathsf{t}}
\left(\B{G}_1^\dag\B{G}_1\right)^2
\B{G}_2^{\dag}\B{G}_2+\np\B{G}_1^{\dag}\B{G}_1\B{G}_2^{\dag}\B{G}_2\right)}}.
\end{align}

From \eqref{sigr}, the received signal at $\Destination{k}$ for
MRC/MRT at the relay is given by
\begin{align}
\label{sigr mrc}
   \yDk \!\!=\! a_{\mathsf{mrc}}\sqrt{P_\mathsf{t}} \B{g}_{2_k}^{\dag} \B{G}_2 \B{G}_1^{\dagger} \B{g}_{1_k} x_k
    &\!+\!a_{\mathsf{mrc}}\sqrt{P_\mathsf{t}}\!\!\!\!\sum_{i=1,i\neq k}^{K}\!\!\!
    \B{g}_{2_k}^{\dag} \B{G}_2 \B{G}_1^{\dagger} \B{g}_{1_i} x_i
        \nonumber
    \\
    &\hspace{-0.5cm}+ a_{\mathsf{mrc}}\B{g}_{2_k}^{\dag} \B{G}_2 \B{G}_1^{\dagger} \nRe + \nDk.
\end{align}
Hence, the e2e SINR can be expressed as
\begin{align}
\label{mrc_sinr1} \gamma^{\mathsf{mrc}}_k
=
\frac{P_{\mathsf{t}}\frac{|\B{g}_{2_k}^{\dag}
\B{G}_2\B{G}_1^{\dag} \B{g}_{1_k}|^2}{\|\B{g}_{2_k}^{\dag}
\B{G}_2\B{G}_1^{\dag}\|^2}} {P_{\mathsf{t}} \sum\limits_{i=1,i\neq
k}^{K} \frac{|\B{g}_{2_k}^{\dag} \B{G}_2\B{G}_1^{\dag}
\B{g}_{1_i}|^2}{\|\B{g}_{2_k}^{\dag} \B{G}_2\B{G}_1^{\dag}\|^2}
    +\np + \frac{\np}{a_{\mathsf{mrc}}^2\|\B{g}_{2_k}^{\dag} \B{G}_2\B{G}_1^{\dag}\|^2}
    }.
\end{align}

\subsection{ZF at the Relay}
We now consider the use of ZF receivers and precoders at the
relay. With ZF processing, the transformation matrix can be
expressed as $\B{W}=a_{\mathsf{zf}}\B{G}_2
\left(\B{G}_2^{\dag}\B{G}_2\right)^{-1}
\left(\B{G}_1^{\dag}\B{G}_1\right)^{-1} \B{G}_1^{\dag} $, where
$a_{\mathsf{zf}}$ is chosen to satisfy the power constraint at the
relay\footnote{Reference \cite[Sec. IV]{Louie} also considers ZF at the relay and employs a \emph{fixed gain} for long term power normalization. In contrast, we consider a variable gain in \eqref{zfg}. Since for receive filtering/transmit ZF precoding at the relay, \emph{instantaneous} channel information must be used and a fixed gain can result in high peak-to-average power ratio signals, the gain in \eqref{zfg} is a good practical choice when implementing ZF at the relay.}, i.e.,
\begin{align}
\label{zfg}
a_{\mathsf{zf}} =
\sqrt{\frac{P_{\mathsf{r}}}{\mathsf{Tr}\left(P_{\mathsf{t}}\left(\B{G}_2^\dag\B{G}_2\right)^{-1}+\np\left(\B{G}_2^\dag\B{G}_2\right)^{-1}\left(\B{G}_1^\dag\B{G}_1\right)^{-1}\right)}}.
\end{align}
For this case, we have
\begin{align}
    \B{g}_{2_k}^{\dag} \B{W} \B{g}_{1_i}
    =
        a_{\mathsf{zf}}\delta_{ki},
\end{align}
where $\delta_{ki}=1$ when $k=i$ and $0$ otherwise. Therefore, from \eqref{sigr}, we can write the received signal at $D_k$ as
\begin{align}
\label{ZFrecv} y_{D_k} = a_{\mathsf{zf}}\sqrt{P_{\mathsf{t}}}
x_{k} +
a_{\mathsf{zf}}\left[\left(\B{G}_1^{\dag}\B{G}_1\right)^{-1}\B{G}_1^{\dag}\right]_{{k}}\nRe
+ n_{D_k},
\end{align}
where $\left[\B{A}\right]_{k}$ is the $k$-th row of the matrix
$\B{A}$.

Now we can express the e2e signal-to-noise ratio (SNR) as
\begin{align}
\label{zf_sinr1}
\gamma^{\mathsf{zf}}_k = \frac{a^2_{\mathsf{zf}}P_{\mathsf{t}}}{a^2_{\mathsf{zf}}\left[\left(\B{G}_1^{\dag}\B{G}_1\right)^{-1}\right]_{{kk}} \np + \np}.
\end{align}

\subsection{Orthogonal Scheme}
For comparison with MRC/MRT and ZF, we also consider a ``\emph{naive scheme}'' that employs orthogonal channel access. Specifically, to completely avoid the inter-user interference, each $S_k-D_k$ pair for $k=1,\ldots,K$ uses $\frac{1}{2K}$ channel resources for communication. At the relay, MRC/MRT is employed as it maximizes the e2e SNR\footnote{In the remainder of the paper, ``MRC/MRT'' is used to refer to the scheme (cf. Section II-A) where $K>1$ destinations are simultaneously served.}. Therefore, the e2e SNR at $D_k$ can be expressed as
\begin{align}
\gamma^{\mathsf{ns}}_k =\frac{\frac{P_{\mathsf{t}}\|\B{g}^{\dag}_{1_k}\|^2}{\np}\frac{P_{\mathsf{r}}\|\B{g}_{2_k}\|^2}{\np}}{\frac{P_{\mathsf{t}}\|\B{g}^{\dag}_{1_k}\|^2}{\np}+\frac{P_{\mathsf{r}}\|\B{g}^{\dag}_{2_k}\|^2}{\np}+1}.
\end{align}

\section{Large $N$ Analysis}
In this section, we further simplify the e2e SINR expressions
\eqref{mrc_sinr1} and \eqref{zf_sinr1} in the very large $N$
regime.  These new expressions illuminate several aspects of the
achievable rate vs. power efficiency performance in the considered
network. Here we assume that $N \gg K$. We further assume that
when $N$ is large, the elements of the channel matrices $\B{G}_1$
and $\B{G}_2$ are still independent. Note that even with very large $N$, the
physical size of the antenna array can be small. For example, at
$2.6$ GHz, a cylindrical array with $128$ antennas and $\lambda/2$ antenna spacing occupies only a
physical size of $28~\text{cm} \times 29$ cm and even with this array, the antennas experience nearly independent fading \cite{RPLLMET:11:SPM,Gao}.

The achievable ergodic sum rate of the system is given
by\footnote{The \emph{exact} analysis of
$C_{\mathsf{sum}}^{\mathsf{mrc}}$ and
$C_{\mathsf{sum}}^{\mathsf{zf}}$ for \emph{arbitrary} $N$ is not a
mathematically tractable problem since the required probability
density functions (p.d.f.s) of $\gamma^{\mathsf{mrc}}_{k}$ and
$\gamma^\mathsf{zf}_{k}$ do not readily permit mathematical
manipulation. A closed-form expression
$C_{\mathsf{sum}}^{\mathsf{ns}}$ can be derived, but not reported
since our main focus in this paper is to analyze the impact of the
very large array, where $N \gg K$.}

\begin{align}
\label{sum_rate}
C_{\mathsf{sum}}^{\star} =
\E\left\{\sum^{K}_{k=1}\frac{1}{2\alpha_{\mathsf{f}}}\log_{2}\left(1+\gamma^{\star}_{k}\right)\right\},
\end{align}
where $\star=\{\mathsf{mrc},\mathsf{zf}, \mathsf{ns}\}$ refers to MRC/MRT, ZF and naive schemes and the
pre-log factor $\frac{1}{2}$ is due to the half-duplex relaying. For MRC/MRT and ZF: $\alpha_{\mathsf{f}}=1$ and for the naive scheme $\alpha_{\mathsf{f}}=K$.

In the following analysis, we will consider three
cases: namely, \emph{Case I)} \emph{Fixed} $N P_{\mathsf{t}}, N
\rightarrow \infty$; \emph{Case II)} \emph{Fixed} $N
P_{\mathsf{r}}, N \rightarrow \infty$; \emph{Case III)} \emph{Fixed}
$N P_{\mathsf{t}}$, \emph{Fixed} $N P_{\mathsf{r}}, N \rightarrow
\infty$.

\subsection{MRC/MRT at the Relay}
\emph{Case I)}: If $P_{\mathsf{t}}=\frac{E_{\mathsf{t}}}{N}$ where
$E_{\mathsf{t}}$ is fixed, then from \eqref{sigr mrc} we have
\begin{align}
\label{largeN_eq1}
    \frac{\yDk}{\sqrt{N}}
    &\!=\!
\frac{a_{\mathsf{mrc}}
\sqrt{E_{\mathsf{t}}}\B{g}^{\dag}_{2_k}\B{G}_2\B{G}_1^{\dag}\B{g}_{1_k}x_k}{N}+\!\!\!\!\!\!\!\sum^{K}_{i=1,
i \neq k}\!\!\!\!\!
\frac{a_{\mathsf{mrc}}\sqrt{E_{\mathsf{t}}}\B{g}^{\dag}_{2_k}\B{G}_2\B{G}_1^{\dag}\B{g}_{1_i}x_i}{N}\nonumber\\
&+\frac{a_{\mathsf{mrc}}\B{g}^{\dag}_{2_k}\B{G}_2\B{G}_1^{\dag}\nRe}{\sqrt{N}}+\frac{\nDk}{\sqrt{N}}.
\end{align}
In the very large $N$ regime, we apply the law of large numbers given by~\cite{cramer}
\begin{align}
\label{largeN_eq5}
  \frac{\B{g}^{\dag}_{2_k}\B{g}_{2_i}}{N} \mathop  \to \limits^{a.s.}_{N\to\infty} \left\{
  \begin{array}{l l}
    0 & \quad i \neq k \\
    1 & \quad i = k \\
  \end{array} \right.
\end{align}
and note that
\begin{align}
N a_{\mathsf{mrc}}
    & =
\sqrt{\frac{P_{\mathsf{r}}}{\mathsf{Tr}\left(
    \frac{E_{\mathsf{t}}}{N^3}\left(\B{G}_1^\dag\B{G}_1\right)^2
\B{G}_2^{\dag}\B{G}_2+\np\frac{1}{N^2}\B{G}_1^{\dag}\B{G}_1\B{G}_2^{\dag}\B{G}_2
    \right)}}\nonumber
\\ & \mathop  \to \limits^{a.s.}_{N\to\infty}
\sqrt{\frac{P_{\mathsf{r}}}{\mathsf{Tr}\left(E_{\mathsf{t}}\B{D}_1^2\B{D}_2+\np\B{D}_1\B{D}_2\right)}}.
\end{align}
Now re-expressing the first term in \eqref{largeN_eq1} as
$$a_{\mathsf{mrc}}\sqrt{E_{\mathsf{t}}}\:\frac{\B{g}^{\dag}_{2_k}\B{G}_2\B{G}_1^{\dag}\B{g}_{1_k}}{N}
= N
a_{\mathsf{mrc}}\sqrt{E_{\mathsf{t}}}\sum^{K}_{i=1}\frac{\B{g}^{\dag}_{2_k}\B{g}_{2_k}\B{g}^{\dag}_{1_k}\B{g}_{1_k}}{N^2}.$$
Therefore, when $N$ tends to infinity, we have
\begin{align}
\label{largeN_eq2}
a_{\mathsf{mrc}}\sqrt{E_{\mathsf{t}}}\:\frac{\B{g}^{\dag}_{2_k}\B{G}_2\B{G}_1^{\dag}\B{g}_{1_k}}{N}
\mathop  \to \limits^{a.s.}
\sqrt{\!\!\frac{P_{\mathsf{r}}E_{\mathsf{t}}}{\mathsf{Tr}\!\left(\!E_{\mathsf{t}}\B{D}_1^2\B{D}_2\!+\!\np\B{D}_1\B{D}_2\!\right)}}
\eta_{1k}\eta_{2k}.
\end{align}
Similarly, for $i \neq k$, re-expressing the second term in \eqref{largeN_eq1} as
$$a_{\mathsf{mrc}}\sqrt{E_{\mathsf{t}}}\:\frac{\B{g}^{\dag}_{2_k}\B{G}_2\B{G}_1^{\dag}\B{g}_{1_i}}{N}
= N
a_{\mathsf{mrc}}\sqrt{E_{\mathsf{t}}}\sum^{K}_{j=1}\frac{\B{g}^{\dag}_{2_k}\B{g}_{2_j}\B{g}_{1_j}^{\dag}\B{g}_{1_i}}{N^2},$$
we obtain
\begin{align}
\label{largeN_eq3}
a\sqrt{E_{\mathsf{t}}}\:\frac{\B{g}^{\dag}_{2_k}\B{G}_2\B{G}_1^{\dag}\B{g}_{1_i}}{N^2} \rightarrow 0.
\end{align}
Note that the third term in \eqref{largeN_eq1} can be written as
$$a_{\mathsf{mrc}}\frac{\B{g}^{\dag}_{2_k}\B{G}_2\B{G}_1^{\dag}\nRe}{\sqrt{N}}
= N
a_{\mathsf{mrc}}\sum^{K}_{i=1}\frac{\B{g}^{\dag}_{2_k}\B{g}_{2_i}\B{g}^{\dag}_{1_i}\nRe}{N\sqrt{N}}.$$
We apply the law of large numbers and the Lindeberg-L\'{e}vy central
limit theorem and obtain\footnote{
    Lindeberg-L\'{e}vy central limit
theorem: Let $\B{p}$ and $\B{q}$ be $n \times 1$ vectors
  whose elements are i.i.d. random variables with zero mean and variances of $\sigma_{p}^2$ and $\sigma_{q}^2$, respectively. Then
  $   \frac{1}{\sqrt{n}} \B{p}^H \B{q} \mathop  \to \limits^d \CG{0}{\sigma_p^2 \sigma_q^2}, ~ \text{as} ~
    n \rightarrow \infty$.}
\begin{align}
\label{largeN_eq7}
a_{\mathsf{mrc}}\frac{\B{g}^{\dag}_{2_k}\B{G}_2\B{G}_1^{\dag}\B{g}_{1_i}}{\sqrt{N}}
\mathop  \to \limits^{d}
\sqrt{\frac{P_{\mathsf{r}}}{\mathsf{Tr}\left(E_{\mathsf{t}}\B{D}_1^2\B{D}_2+\np\B{D}_1\B{D}_2\right)}}
\eta_{2k}\widetilde{n}_R,
\end{align}
where $\widetilde{n}_R \sim
\mathcal{CN}\left(0,\eta_{1k}\np\right)$. Substituting \eqref{largeN_eq2}, \eqref{largeN_eq3} and \eqref{largeN_eq7} into
\eqref{largeN_eq1} and since $\frac{1}{\sqrt{N}}n_{D_k}
\rightarrow 0$ we have
\begin{align}
\label{dcaseA}
\frac{\yDk}{\sqrt{N}} & \rightarrow
\sqrt{\frac{P_{\mathsf{r}}E_{\mathsf{t}}}{\mathsf{Tr}\left(E_{\mathsf{t}}\B{D}_1^2\B{D}_2+\np\B{D}_1\B{D}_2\right)}}\eta_{1k}\eta_{2k}x_k
\nonumber
\\
&+\sqrt{\frac{P_{\mathsf{r}}}{\mathsf{Tr}\left(E_{\mathsf{t}}\B{D}_1^2\B{D}_2+\np\B{D}_1\B{D}_2\right)}}\eta_{2k}\widetilde{n}_R.
\end{align}
Now from \eqref{dcaseA} we obtain
\begin{align}\label{eq mrc inf1}
\gamma^{\mathsf{mrc}}_k \rightarrow \frac{E_{\mathsf{t}} \eta_{1k}}{\np}, ~
\text{as} ~ N \to \infty.
\end{align}

For $K=1$, it is clear that we can reduce the transmit power by a
factor of $1/N$ with no reduction in performance due to the array
gain. But here we consider multiple sources, and the above result
implies that by using a large number of relay antennas, we can
still obtain the same array gain as in the case of single source.
Interestingly, from \eqref{eq mrc inf1} we can see that when $N$
grows large and $N P_{\mathsf{t}}$ is fixed, the e2e SNR does not
depend on the transmit power at the relay and the large-scale fading of the second hop. This is due to the fact that, with MRT precoding at the relay in the
second phase and $P_{\mathsf{r}}$ is fixed, as $N$ goes to
infinity, the effect of inter-user interference and  noise at $D_k$ disappears. Finally, the sum rate follows directly by substituting \eqref{eq mrc inf1} into \eqref{sum_rate}.

\emph{Case II)}: If $P_{\mathsf{r}}=\frac{E_{\mathsf{r}}}{N}$ where
$E_{\mathsf{r}}$ is fixed, we observe that
\begin{align}
N^2 a_{\mathsf{mrc}}
    & \mathop  \to \limits^{a.s.}_{N\to\infty}
\sqrt{\frac{E_{\mathsf{r}}}{\mathsf{Tr}\left(P_{\mathsf{t}}\B{D}_1^2\B{D}_2\right)}}.
\end{align}
Therefore when $N$ grows without bound, the first term in \eqref{sigr
mrc} tends to
\begin{align}
a_{\mathsf{mrc}}\sqrt{P_{\mathsf{t}}}\B{g}^{\dag}_{2_k}\B{G}_2\B{G}_1^{\dag}\B{g}_{1_k}
\mathop  \to \limits^{a.s.}
\sqrt{\frac{E_{\mathsf{r}}}{\mathsf{Tr}\left(\B{D}_1^2\B{D}_2\right)}}\eta_{1k}\eta_{2k}.
\end{align}
Similarly, when $N$ goes to infinity, the second and third terms
in \eqref{sigr mrc} converge almost sure to $0$. Therefore,
\begin{align} \label{B inf}
\yDk & \mathop  \to \limits^{a.s.}
\sqrt{\frac{E_{\mathsf{r}}}{\mathsf{Tr}\left(\B{D}_1^2\B{D}_2\right)}}\eta_{1k}\eta_{2k}
x_k + n_{D_k}, ~ \text{as} ~ N\to \infty.
\end{align}
From \eqref{B inf}, as $N \to \infty$, we have
\begin{align}
\gamma^{\mathsf{mrc}}_k \rightarrow
\frac{E_{\mathsf{r}}\eta^2_{1k}\eta^2_{2k}}{\mathsf{Tr}\left(\B{D}_1^2\B{D}_2\right)\np}.
\end{align}

The above result shows that the transmit power at relay can be made inversely proportional to the number of relay antennas without compromising the quality-of-service. Note that in the special case where $\eta_{1k}=\eta_1$ and $\eta_{2k}=\eta_2$ for $k=1,\ldots,K$, we have $\gamma^{\mathsf{mrc}}_k \rightarrow \frac{1}{K}\frac{E_{\mathsf{r}}\eta_2}{\np}$ which does not depend on the transmit power of each source and the channel quality of the first hop.

\emph{Case III)}: If $P_{\mathsf{t}}=\frac{E_{\mathsf{t}}}{N}$ and
$P_{\mathsf{r}}=\frac{E_{\mathsf{r}}}{N}$ where $E_{\mathsf{t}}$
and $E_{\mathsf{r}}$ are fixed, using a similar approach as above
we can show that
\begin{align}
\yDk
    &\mathop  \to \limits^{d}_{n\to\infty}
\sqrt{\frac{E_{\mathsf{t}}E_{\mathsf{r}}}{\mathsf{Tr}\left(E_{\mathsf{t}}\B{D}_1^2\B{D}_2+\np\B{D}_1\B{D}_2\right)}}\eta_{1k}\eta_{2k}x_k
\nonumber\\
&+\sqrt{\frac{E_{\mathsf{r}}}{\mathsf{Tr}\left(E_{\mathsf{t}}\B{D}_1^2\B{D}_2+\np\B{D}_1\B{D}_2\right)}}\eta_{2k}\widetilde{n}_R+\nDk.
\end{align}
Therefore,
\begin{align}
\gamma^{\mathsf{mrc}}_k \rightarrow \frac{\frac{E_{\mathsf{t}}
\eta_{1k}}{\np}}{1+\frac{\mathsf{Tr}\left(E_{\mathsf{t}}\B{D}_1^2\B{D}_2+\np\B{D}_1\B{D}_2\right)}{E_{\mathsf{r}}\eta_{1k}\eta^2_{2k}}},
~ \text{as} ~ N \to\infty.
\end{align}

Interestingly, by using large antenna arrays, we can scale down
both transmit powers of source and relay nodes by a factor of
$1/N$ with no reduction in performance. Furthermore, we can see
that when $E_{\mathsf{t}}\to\infty$, the above SINR coincides with
the result for the case $P_{\mathsf{r}}$ is fixed, and when
$E_{\mathsf{r}}\to\infty$, the above SINR coincides with the
result for the case $P_{\mathsf{t}}$ is fixed.

In the special case where $\eta_{1k}=\eta_1$ and
$\eta_{2k}=\eta_2$ for $k=1,\ldots,K$; we have $\gamma^{\mathsf{mrc}}_k \rightarrow
\frac{\frac{E_{\mathsf{t}}
\eta_{1}}{\np}}{1+\frac{K}{\frac{E_{\mathsf{r}}
\eta_{2}}{\np}}\left(1+\frac{E_{\mathsf{t}}
\eta_{1}}{\np}\right)}$.

\subsection{ZF at the Relay}
By following a similar derivation as in the case of MRC/MRT, we
can obtain the same power scaling law as follows.

\emph{Case I)}: In the very large $N$ regime we first note that
\begin{align}
\frac{a_{\mathsf{zf}}}{N}
     \mathop  \to \limits^{a.s.}
\sqrt{\frac{P_{\mathsf{r}}}{\mathsf{Tr} \left(
E_{\mathsf{t}}\B{D}_2^{-1}+\np
\left(\B{D}_1\B{D}_2\right)^{-1}\right)}}.
\end{align}
Then from \eqref{ZFrecv} we have
\begin{align}
\frac{y_{D_k}}{\sqrt{N}} &
    \mathop  \to \limits^{d}_{N\to\infty}
    \sqrt{\frac{P_{\mathsf{r}}}{\mathsf{Tr} \left(
E_{\mathsf{t}}\B{D}_2^{-1}+\np
\left(\B{D}_1\B{D}_2\right)^{-1}\right)}} x_k
    \nonumber\\
&\hspace{-0.5cm}+\sqrt{\frac{P_{\mathsf{r}}}{\mathsf{Tr}\left(E_{\mathsf{t}}\B{D}_2^{-1}+\np
\left(\B{D}_1\B{D}_2\right)^{-1}\right)\eta^2_{1k}}}\widetilde{n}_{R}+\frac{n_{D_k}}{\sqrt{N}},
\end{align}
where $\widetilde{n}_{R} \sim \mathcal{CN}\left(0,
\eta_{1k}\np\right)$. Hence, when $N$ grows without bound, we
obtain
\begin{align}
\gamma^{\mathsf{zf}}_{k} \mathop  \to \limits^{a.s.}
\frac{E_{\mathsf{t}}\eta_{1k}}{\np}.
\end{align}

\emph{Case II)}: In this case for very large $N$, $a_{\mathsf{zf}}$
tends to
\begin{align}
a_{\mathsf{zf}} \mathop  \to \limits^{a.s.}
\sqrt{\frac{E_{\mathsf{r}}}{\mathsf{Tr}\left(P_{\mathsf{t}}\B{D}_2^{-1}\right)}}.
\end{align}
Therefore, we can write \eqref{ZFrecv} as
\begin{align}
y_{D_k} & \rightarrow
\sqrt{\frac{E_{\mathsf{r}}}{\mathsf{Tr}\left(P_{\mathsf{t}}\B{D}_2^{-1}\right)}}\sqrt{P_{\mathsf{t}}}x_k
+n_{D_k}, ~ \text {as} ~ N\to\infty.
\end{align}
which leads to
\begin{align}
\gamma^{\mathsf{zf}}_k \rightarrow
\frac{E_{\mathsf{r}}}{\mathsf{Tr}\left(\B{D}_2^{-1}\right)\np}, ~
\text {as} ~ N\to\infty.
\end{align}

In the special case when $\eta_{2k}=\eta_2$ for $k=1,\ldots,K$, we
have $\gamma_k \rightarrow
\frac{1}{K}\frac{E_{\mathsf{r}}\eta_2}{\np}$.

\emph{Case III)}: If $P_{\mathsf{t}}=\frac{E_{\mathsf{t}}}{N}$ and
$P_{\mathsf{r}}=\frac{E_{\mathsf{r}}}{N}$ where $E_{\mathsf{t}}$
and $E_{\mathsf{r}}$ are fixed, when $N$ tends to infinity, we
obtain
\begin{align}
\frac{a_{\mathsf{zf}}}{\sqrt{N}}
    \mathop  \to \limits^{a.s.}
\sqrt{\frac{E_{\mathsf{r}}}{\mathsf{Tr}\left(E_{\mathsf{t}}\B{D}_2^{-1}+\np\left(\B{D}_1\B{D}_2\right)^{-1}\right)}}.
\end{align}
Therefore,
\begin{align}
y_{D_k}
    &\mathop  \to \limits^{d}
     \sqrt{\frac{E_{\mathsf{t}}E_{\mathsf{r}}}{\mathsf{Tr}\left(E_{\mathsf{t}}\B{D}_2^{-1}+\np\left(\B{D}_1\B{D}_2\right)^{-1}\right)}}x_k
     \\\nonumber
&+\sqrt{\frac{E_{\mathsf{r}}}{\mathsf{Tr}\left(E_{\mathsf{t}}\B{D}_2^{-1}+\np\left(\B{D}_1\B{D}_2\right)^{-1}\right)\eta^2_{1k}}}\widetilde{n}_R+n_{D_k},
\end{align}
where $\widetilde{n}_{R} \sim \mathcal{CN}\left(0,
\eta_{1k}\np\right)$. Hence, when $N$ goes to infinity, we obtain
\begin{align}
\gamma^{\mathsf{zf}}_k \rightarrow
\frac{\frac{E_{\mathsf{t}}\eta_{1k}}{\np}}{1+\frac{\mathsf{Tr}\left(E_{\mathsf{t}}\B{D}_2^{-1}+\np\left(\B{D}_1\B{D}_2\right)^{-1}\right)\eta_{1k}}{E_{\mathsf{r}}}}.
\end{align}

In the special case where $\eta_{1k}=\eta_1$ and
$\eta_{2k}=\eta_2$ for $k=1,\ldots,K$, we have $\gamma^{\mathsf{zf}}_k \rightarrow
\frac{\frac{E_{\mathsf{t}}
\eta_{1}}{\np}}{1+\frac{K}{\frac{E_{\mathsf{r}}
\eta_{2}}{\np}}\left(1+\frac{E_{\mathsf{t}}
\eta_{1}}{\np}\right)}$.

\noindent{\emph{Remark 1}}: In \emph{Case I)} the
asymptotic rate $(N \rightarrow \infty)$ of $D_k$ with MRC/MRT and
ZF processing is same and given by
$C^{\mathsf{mrc}}_{k}=C^{\mathsf{zf}}_{k}=\frac{1}{2}\log_2\left(1+\frac{E_{\mathsf{t}}\eta_{1k}}{\np}\right)$.

\noindent{\emph{Remark 2}}: In \emph{Case II)} the
asymptotic rates at $D_k$ with MRC/MRT and ZF,
$C^{\mathsf{mrc}}_{k} \lesseqgtr C^{\mathsf{zf}}_{k}$, is
determined by
\begin{align}
\frac{1}{\eta^2_{1k}\eta^2_{2k}}\sum^{K}_{i \neq k} \eta^2_{1i}\eta_{2i} \gtreqless \sum^{K}_{i \neq k} \frac{1}{\eta_{2i}}.
\end{align}

\noindent{\emph{Remark 3}}: In \emph{Case III)} the
asymptotic rates at the $k$th destination with MRC/MRT and ZF,
$C^{\mathsf{mrc}}_{k} \lesseqgtr C^{\mathsf{zf}}_{k}$, is
determined by
\begin{align}
\frac{1}{\eta^2_{1k}\eta^2_{2k}}\sum^{K}_{i \neq k} \eta_{1i}\eta_{2i}\left(1+\frac{E_{\mathsf{t}}\eta_{1i}}{\np}\right) \gtreqless \sum^{K}_{i \neq k}\frac{1}{\eta_{1i}\eta_{2i}}\left(1+\frac{E_{\mathsf{t}}\eta_{1i}}{\np}\right).
\end{align}

\subsection{Orthogonal Scheme}
We now consider the asymptotic sum rate of the orthogonal scheme. In \emph{Case I)}: It is easy to show that $\gamma^{\mathsf{ns}}_k \rightarrow \frac{E_{\mathsf{t}}\eta_{1k}}{\np}$ and $C^{\mathsf{ns}}_{\mathsf{sum}} = \frac{1}{2K}\sum^K_{i=1}\log_2\left(1+\frac{E_{\mathsf{t}}\eta_{1k}}{\np}\right)$.
In \emph{Case II)}: $\gamma^{\mathsf{ns}}_k \rightarrow \frac{E_{\mathsf{r}}\eta_{2k}}{\np}$ and
$C^{\mathsf{ns}}_{\mathsf{sum}} = \frac{1}{2K}\sum^K_{i=1}\log_2\left(1+\frac{E_{\mathsf{r}}\eta_{2k}}{\np}\right)
$ and in \emph{Case III)}: We have $\gamma^{\mathsf{ns}}_k \rightarrow \frac{\frac{E_{\mathsf{t}}\eta_{1k}}{\np}\frac{E_{\mathsf{r}}\eta_{2k}}{\np}}{\frac{E_{\mathsf{t}}\eta_{1k}}{\np}+\frac{E_{\mathsf{r}}\eta_{2k}}{\np}+1}$ and
\begin{align}
C^{\mathsf{ns}}_{\mathsf{sum}} = \frac{1}{2K}\sum^K_{i=1}\log_2\left(1+\frac{\frac{E_{\mathsf{t}}\eta_{1k}}{\np}\frac{E_{\mathsf{r}}\eta_{2k}}{\np}}{\frac{E_{\mathsf{t}}\eta_{1k}}{\np}+\frac{E_{\mathsf{r}}\eta_{2k}}{\np}+1}\right).
\end{align}

\begin{figure}[t]
\centering
\includegraphics[width=0.96\linewidth]{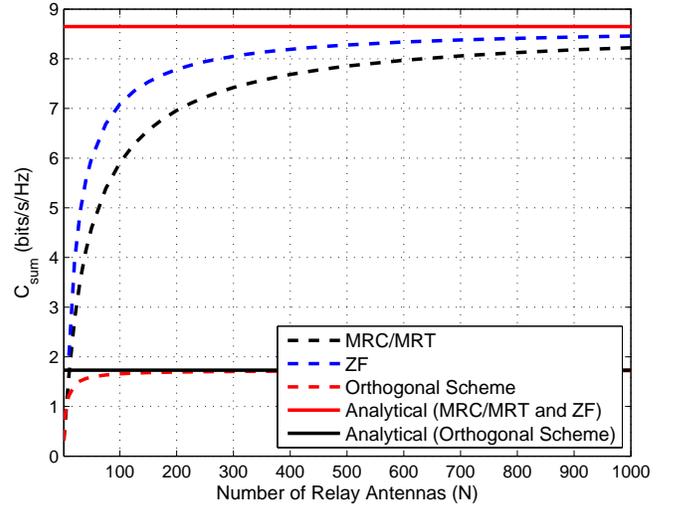}\\
\caption{\emph{Case I)}: Sum rate vs. the number of relay antennas. $E_{\mathsf{t}} = 10$ dB, $P_{\mathsf{r}}=1$ and $K=5$ users are served. $\B{D}_{1}=\B{D}_2=\B{I}_5$.}\label{Fig2}
\end{figure}

\begin{figure}[t]
\centering
\includegraphics[width=0.96\linewidth]{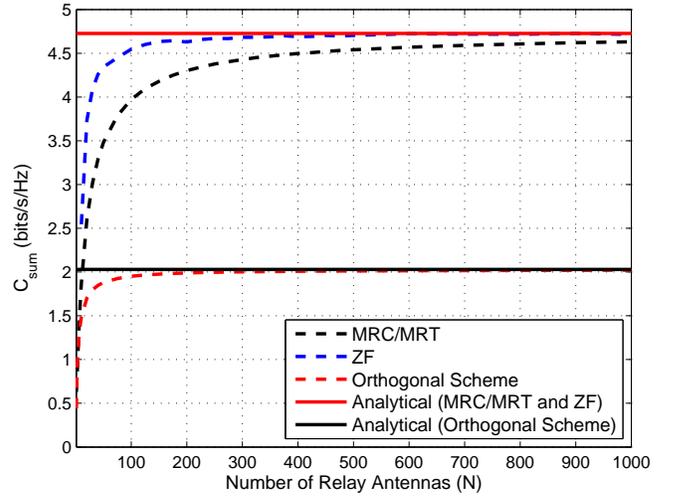}\\
\caption{\emph{Case II)}: Sum rate vs. the number of relay
antennas. $P_{\mathsf{t}} = 1, E_{\mathsf{r}}=10$ dB and $K=5$ users are served. $\B{D}_{1}=\B{D}_2=\B{I}_5$.}\label{Fig3}
\end{figure}

\section{Numerical Results}
The sum rates achieved by the multi-pair relay system are evaluated through simulations and compared with our asymptotic analytical results. Without loss of generality, $\np=1$ is assumed.

Fig. 2 shows the simulated sum rate vs. the number of relay antennas and the presented analytical asymptotic results for $\emph{Case I)}$. Clearly, as the number of antennas increases, the sum rates of MRC/MRC, ZF and the naive schemes approaches the corresponding constant values predicted by our analysis. Interestingly, the sum rate curve of ZF has a sharper knee than the MRC/MRT counterpart on the way to the same asymptotic constant and the achieved sum rate is $8.65$ bits/s/Hz. Though it is not explicitly seen in Fig. 2, for small number of antennas ($N\leq 6$), the naive scheme exhibits a better sum rate than MRC/MRT as expected. However, as $N$ increases, the sum rate offered by the naive scheme rapidly saturates while the sum rates of the MRC/MRT and ZF schemes show a rapid improvement. This is because with a limited number of antennas, interference cannot be significantly reduced and thus lowers the sum rate of MRC/MRT. But when $N$ grows large, the random channel vectors between sources/destinations and relay become pairwise orthogonal and hence, the interference is canceled out. At the same time, we gain from simultaneously serving $K$ source-destination pairs in the same time-frequency resource.

Figs. 3 and 4 show results for the second and third power scaling laws; $\emph{Case II)}$ and $\emph{Case III)}$. Both MRC/MRT and ZF achieves the same sum rate of $4.73$ and $3.36$ bits/s/Hz in $\emph{Case II)}$ and $\emph{Case III)}$, respectively. Moreover, similar trends in results as in Fig. 2 can be observed.

The rates achieved by individual destinations are illustrated in Fig. 5 for slow-fading coefficients; $\eta_{11}=2, \eta_{12}=2,\eta_{13}=2$ and $\eta_{21}=1, \eta_{22}=3,\eta_{23}=3$ and \emph{Case II)}. Results in Fig. 5 confirm that the sum rate of MRC/MRT can be higher than the sum rate of ZF depending
on the slow fading parameters. $C^{\mathsf{mrc}}_{\mathsf{sum}}=8.98$ and $C^{\mathsf{zf}}_{\mathsf{sum}}=8.90$. Recall that $C^{\mathsf{mrc}}_{\mathsf{sum}}<C^{\mathsf{zf}}_{\mathsf{sum}}$ in the example of Fig. 3. Interestingly, when $N \leq 350$, all three users in the ZF system achieve a higher rate than the users in the MRC system. However, when $N$ is very large, two users of the MRC achieve a higher rate than the ZF users.

\begin{figure}[t]
\centering
\includegraphics[width=0.95\linewidth]{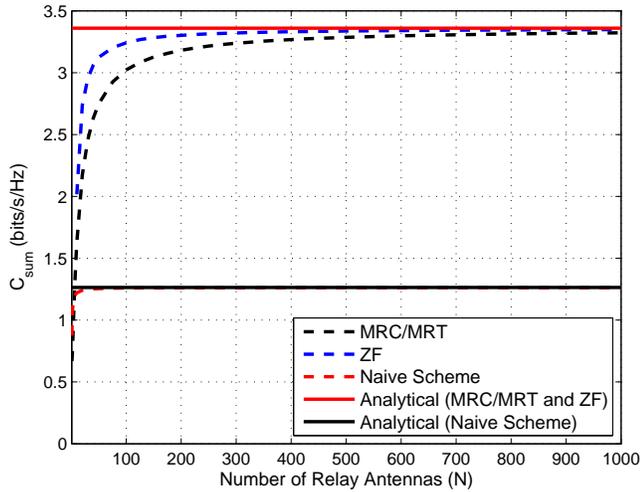}\\
\caption{\emph{Case III)}: Sum rate vs. the number of relay
antennas. $P_{\mathsf{t}}=10$ dB, $E_{\mathsf{r}}=10$ dB and $K=5$ users are served. $\B{D}_{1}=\B{D}_2=\B{I}_5$.}\label{Fig4}
\end{figure}

\begin{figure}[t]
\centering
\includegraphics[width=0.94\linewidth]{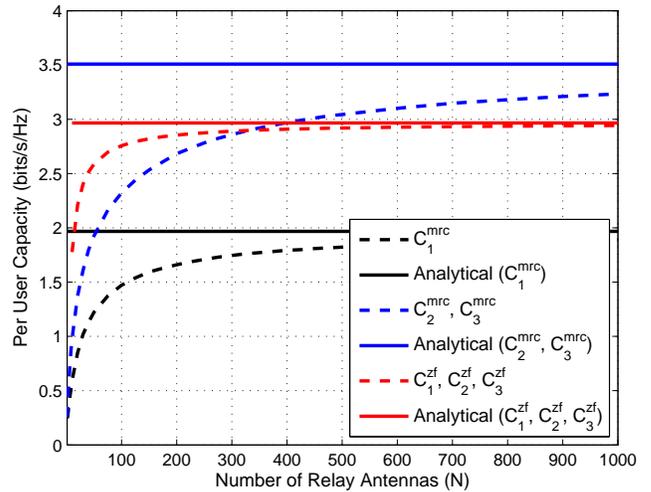}\\
\caption{\emph{Case II)}: Per user rate vs. the number of relay antennas. $K=3$ are served.}\label{Fig5}
\end{figure}

\section{Conclusion}
We have shown that relay systems can benefit significantly from the use of very large antenna arrays. The offered sum rates of a multi-pair relay system was investigated for three different power scaling laws. At the relay, MRC/MRT and ZF processing was
considered. We derived asymptotic sum rate results and confirmed their accuracy using computer simulations. Several insights were extracted using the analysis to illuminate the comparative performances between MRC/MRT and ZF. For example, the asymptotic
achievable rates of MRC/MRT and ZF are the same if we scale the power at the sources.

\end{document}